\documentclass[aps,pra,twocolumn,superscriptaddress,showpacs,preprintnumbers,amsmath,amssymb,floatfix]{revtex4-1}
\usepackage{amsmath}
\usepackage{graphicx}
\usepackage{dcolumn}
\usepackage{color}

\begin{document}

\title{Bound vortex states and exotic lattices in multi-component Bose-Einstein condensates: The role of vortex-vortex interaction}

\author{Davi S. Dantas}\email{davidantas@fisica.ufc.br}
\affiliation{Universidade Federal do Cear\'a, Departamento de
F\'{\i}sica Caixa Postal 6030, 60455-760 Fortaleza, Cear\'a, Brazil}
\author{Aristeu R. P. Lima}\email{aristeu@unilab.edu.br}
\affiliation{Universidade da Integra\c{c}\~{a}o Internacional da Lusofonia Afro-Brasileira, Campus dos Palmares, 
62785-000 Acarape, Cear\'{a}, Brazil}
\author{A. Chaves}\email{andrey@fisica.ufc.br}
\author{C. A. S. Almeida}
\author{G. A. Farias}
\affiliation{Universidade Federal do Cear\'a, Departamento de
F\'{\i}sica Caixa Postal 6030, 60455-760 Fortaleza, Cear\'a, Brazil}
\author{M. V. Milo\v{s}evi\'c}\email{milorad.milosevic@uantwerpen.be}
\affiliation{Departement Fysica, Universiteit Antwerpen, Groenenborgerlaan 171, B-2020, Belgium}

\begin{abstract}
We numerically study the vortex-vortex interaction in multi-component homogeneous Bose-Einstein condensates within the realm of the Gross-Pitaevskii theory. We provide strong evidences that pairwise vortex interaction captures the underlying mechanisms which determine the geometric configuration of the vortices, such as different lattices in many-vortex states, as well as the bound vortex states with two (dimer) or three (trimer) vortices. Specifically, we discuss and apply our theoretical approach to investigate intra- and inter-component vortex-vortex interactions in two- and three-component Bose-Einstein condensates, thereby shedding light on the formation of the exotic vortex configurations. These results correlate with current experimental efforts in multi-component Bose-Einstein condensates, and the understanding of the role of vortex interactions in multiband superconductors.
\end{abstract}


\maketitle

\section{Introduction}

The realization of Bose-Einstein condensates (BECs) has brought about a suitable research environment for investigating general properties of superfluidity and superconductivity with a high degree of control and versatility \cite{anderson_mh_1995,bradley_cc_1995,davis_kb_1995}. This has led vortex states and their dynamics, key concepts for both superfluidity and superconductivity, to the rank of some of the most investigated topics in low temperature physics \cite{fetter_review,cooper_review,yarmchuk}. Remarkably, since the first observation of vortices \cite{MadisonI, MadisonII} and, subsequently, the formation of highly-ordered vortex lattices in BECs \cite{Abo-Shaeer}, much theoretical and experimental effort has been made to push further the understanding of these systems. Specially, recent advances in experimental techniques have led to the realization of condensates with several different types of particle-particle interaction, making this research field even broader. In the spotlight, theoretical results on dipolar \cite{cooper_stripes,zhang_stripes} and spin-orbit-coupled \cite{spielman} BECs exhibit rich ground-state phases, with bubbles, density stripes and various vortex lattice geometries. Besides the already mentioned possibilities, further remarkable physical phenomena to be addressed with ultracold atomic systems include, for example, quantum-fluid turbulence in BECs \cite{Henn}, multicharged vortices \cite{ednilson} and vortex-antivortex lattices in superfluid fermionic gases \cite{Botelho}.

With the recent prospects of producing condensates with a large number of components and different types of interaction \cite{mingwu,aikawa}, vortex-states in multi-component condensates become even more important. Indeed, recent theoretical results on vortex lattice conformations have shown that multi-component Bose-Einstein condensates are far from being just a trivial extension of the single component case. In fact, adding a component brings a diversity of possible configurations never found in a one-component system, such as amorphous conformations, square lattices and bound states, including overlapped vortices, vortex dimers and molecules \cite{Kasamatsu, Kuopanportti, Eto,MCipriani,chao_fei}. Some of these conformations suggest that the interaction between vortices can be nonmonotonic with respect to the inter-vortex distance. Unfortunately, to date, the interaction between vortices in the simplest multi-component BEC is known only in specific limits of scale, by either assuming inter-vortex distances much greater than the healing length \cite{Eto2} or considering the interaction energy near the vortex peak in the Thomas-Fermi regime \cite{Aftalion}. As it turns out, the asymptotic behavior does not account for all conformations found \cite{Kuopanportti,Eto,MCipriani,chao_fei}. Furthermore, the generalization of these analytical approaches to more complex cases with more components or even with a different kind of inter-component particle-particle interaction seems to be highly non-trivial.

Within this context, we investigate in the present paper the origin of these unusual quantized vortex states by focusing on the pairwise vortex-vortex interaction \cite{Chaves1, Chaves2}. We consider homogeneous BECs, which possess translation invariance. On the one hand, this is important in itself, since the key properties of the experimentally more relevant harmonically trapped BECs in the large particle number regime bear close resemblance to those of their homogeneous counterparts \cite{aristeu}. On the other hand, the recent achievement of a condensate in a uniform potential \cite{zoran} enables the experimental verification of our predictions. The experimental progress has already led to the detection of extended phase coherence in a uniform quasi-two-dimensional Bose gas \cite{chomaz}. In our approach, constraints are imposed on the Gross-Pitaevskii (GP) formalism, allowing for fixing the vortices in desired positions for further analysis. In other words, vortices no longer arise naturally from the GP equation itself, but are instead placed manually in the position of interest. This brings about the possibility of calculating the interaction energy between vortices as a function of their distance. Then, by investigating conformations which minimize the corresponding energy, we are able to present a simple physical picture of the underlying phenomena which lead to elsewhere observed vortex states. 

This paper is organized as follows. In Sec. II, we briefly present the main aspects of the theoretical approach, focusing on the Euler-Lagrange equations. Sec. III summons our numerical results and displays a discussion concerning experimental realization in each considered case. While section III.A focuses on two-component BECs with contact interaction, section III.B concerns bound states of vortices in two- and three-component condensates featuring also a coherent coupling of the Rabi type. Our concluding remarks are presented in Sec. IV.

\section{Theoretical Approach}
\label{SecI}
We begin our consideration from the Gross-Pitaevskii energy functional for an $N_{c}$-component homogeneous BEC. When set into rotation above a given critical frequency, superfluids acquire angular momentum in the form of a vortex \cite{butts_rokhsar,linn}. In turn, these can be represented by nodes in the wave-function $\Psi$ \cite{feynman}. Here we adopt a different approach and include vortices directly in the form of a node, disregarding rotation. Therefore, only contributions from the kinetic and interaction energies should enter in the energy functional. Moreover, since vortices will be placed manually in fixed positions, we also do not need to account for the energy provided by the angular momentum of the system, $L_{z}$, which is only necessary to allow the vortex formation in the Gross-Pitaevskii theory for a rotating BEC. Then, in the present case, the energy density functional reads 
\begin{equation}
\label{Eq1}
\begin{aligned} 
&\mathcal{E} = \sum_{\alpha = 1}^{N_{c}} \left\{\frac{\hbar^2}{2m_{\alpha}}|\nabla \Psi_{\alpha}|^2 + \frac{g_{\alpha}}{2}|\Psi_{\alpha}|^4 \right\} + V_{ij}, 
\end{aligned}
\end{equation}
where the components are labeled by the index $\alpha$. Here, the first and second terms, respectively, account for the usual kinetic and contact interaction energies within each component. In addition, the term $V_{ij}$ stands for the inter-component coupling energy density.

In what follows, we will be interested in cases where $V_{ij}$ can be written as
\begin{equation}
\label{Eq2}
V_{ij} = \sum_{i}^{N_{c}}\sum_{j>i}^{N_{c}}g_{ij}|\Psi_{i}|^2|\Psi_{j}|^2-w_{ij}\left(\Psi_{i}\Psi^{*}_{j}+\Psi_{j}\Psi^{*}_{i}\right),
\end{equation}
where the first term stands for the contact interaction and the last one represents the Rabi term. The latter characterizes the internal coherent coupling between the components and has been already experimentally implemented by inducing an external driving field, which allows particles to move from one hyperfine spin state to another \cite{Matthews}. These interaction terms are controlled by the parameters $g_{\alpha}$, $g_{ij}$ and $w_{ij}$, respectively. However, it is convenient to redefine them in order to have dimensionless units of energy and length. Therefore, for a two-dimensional system, we introduce the units $\mathcal{E}_{1}=\hbar^{2}\bar{\rho}_{1}^2/2m_{1}$ for energy density and $\xi=1/\sqrt{\bar{\rho}_{1}}$ for distances. Here, $\bar{\rho}_{1}$ is the average particle density of the first species. With  $E =\mathcal{E}/\mathcal{E}_{1}$ and $\vec{r}=\vec{r}'/\xi$, the energy density can be written in its dimensionless form as
\begin{equation} 
\label{Eq3}
\begin{aligned}
&E= \sum_{\alpha=1}^{N_{c}}\left[M_{1\alpha}|\nabla \psi_{\alpha}|^2 + \frac{\gamma_{\alpha}}{2}|\psi_{\alpha}|^4\right]\\& +\sum_{i}^{N_{c}}\sum_{j>i}^{N_{c}}\left[\gamma_{ij}|\psi_{i}|^2|\psi_{j}|^2-\omega_{ij}\left(\psi_{i}\psi_{j}^{*}+\psi_{j}\psi_{i}^{*}\right)\right],
\end{aligned}
\end{equation}
where $\psi_{\alpha}=\xi\Psi_{\alpha}$ is the dimensionless order parameter. This procedure leads to the definition of the mass ratio $M_{1\alpha} = m_1/m_{\alpha}$, the dimensionless contact interaction strengths ${\gamma_{\alpha}}=g_{\alpha}\bar{\rho}_{1}^2/\mathcal{E}_{1}$ and ${\gamma_{ij}}=g_{ij}\bar{\rho}_{1}^2/\mathcal{E}_1$, and the dimensionless Rabi frequencies  $\omega_{ij}=w_{ij}\bar{\rho}_{1}/\mathcal{E}_1$. The contact parameters can be well controlled in experiments by means of Feshbach resonances and appropriate values should enable the visualization of properties of interest.

In order to measure the interaction potential between two vortices, we consider the total energy as a function of the inter-vortex distance. This is justified because the system is free from external contributions, so that the vortex position only affects the vortex-vortex interaction. Moreover, calculating the system energy as a function of the distance between vortex pairs requires the vortex fixing, which is achieved by fixing the $2\pi$ whirl of the phase of the condensate wave function, thereby introducing a node in it. For one vortex in the $\alpha$-component, for example, the wave function is given as $\psi_{\alpha}(x,y)=\sqrt{N_{\alpha}}e^{in_{k}\theta_{k}}f_{\alpha}(x,y)$, with $N_{\alpha}$ the number of particles in the $\alpha$-component and $n_{k}$ the corresponding winding number, i.e., the quanta of circulation carried by the vortex. In addition, the angle $\theta_{k}$ is defined around the locus of the $k$-th vortex. Correspondingly, for two vortices, we have $\psi_{\alpha}(x,y)=\sqrt{N_{\alpha}}e^{in_{1,\alpha}\theta_{1,\alpha}}e^{in_{2,\alpha}\theta_{2,\alpha}}f_{\alpha}(x,y)$. Since such vortex dimer structure is not circularly symmetric, Cartesian coordinates are an appropriate choice, with $e^{i n_{k,\alpha}\theta_{k,\alpha}}$ is written as
\begin{equation}
\label{Eq4}
e^{i n_{k,\alpha} \theta_{k,\alpha}} = \left(\frac{x_{k,\alpha} + iy_{k,\alpha}}{x_{k,\alpha} -
iy_{k,\alpha}}\right)^{n_{k,\alpha}/2},
\end{equation}
where $\vec{r}_{k,\alpha} = (x_{k,\alpha}, y_{k,\alpha}, 0) $ stands for the in-plane position vector with origin at the center of the vortex $k$. As a consequence of the fixed circular phase around the vortex, singularities in the amplitudes appear naturally from the energy functional minimization.

We remark that the present ansatz is general enough to allow for considering components whose vortices might differ both in position and winding number. This is important in order to obtain the total energy of the system in the presence of the inter-component coupling as a function of the relative distance of the vortices and then verify whether one has found the lowest energy configuration or not. Indeed, in the absence of inter-component coupling $\gamma_{ij}=\omega_{ij}=0$, we have found that the single-vortex state is infinitely degenerate and the respective position of vortices in different components is irrelevant. This supports our statement above that the total energy depends only on the inter-vortex distance. In the presence of an inter-component coupling, however, vortices in different components will rather stay on top of each other (separate away) for attractive (repulsive) effective vortex-vortex interaction potentials.

With the adequate ansatz for the condensate wave function at hand, we minimize the total energy, which is obtained by integrating in space the quantity $E    -\sum\mu_{\alpha}|\psi_{\alpha}|^{2}$, with respect to $f_{\alpha}$. Here, $\mu_{\alpha}$ stands for the chemical potential of the the $\alpha$-component and is introduced to keep the corresponding number of particles constant, which means that we are not taking population transfer between different hyperfine spin states into account. The minimization process is implemented by numerically solving the set of $N_{c}$ Euler-Lagrange equations
\begin{equation}
\label{Eq5}
\begin{aligned}
&M_{1\alpha}K_{\alpha}f_{\alpha} + \mu_{\alpha}f_{\alpha}-\gamma_{\alpha}Nf_{\alpha}^{2}f_{\alpha} \\+&\sum_{i}^{N_{c}}\sum_{j>i}^{N_{c}}\omega_{ij}\Theta_{ij}\left(f_{i}\delta_{\alpha,j}+f_{j}\delta_{\alpha,i}\right)\\-&N\gamma_{ij}\left(f_{j}^{2}f_{i}\delta_{i,\alpha}+f_{i}^{2}f_{j}\delta_{j,\alpha}\right)=0,
\end{aligned}
\end{equation}
where
\begin{equation}
\Theta_{ij}=\cos{\left[\left(n_{1i}\theta_{1i}-n_{1j}\theta_{1j}\right)+\left(n_{2i}\theta_{2i}-n_{2j}\theta_{2j}\right)\right]}.
\end{equation}
At this point, we have chosen the same number of particles for all components by setting $N_{\alpha}=N$. Notice that Eq. (\ref{Eq5}) has the same form of the Gross-Pitaevskii equation and we shall call it constrained GP equation (CGP). One important difference, however, should be emphasized: the term 
\begin{equation}
K_{\alpha}=\nabla^{2}-\left(\overline{X_\alpha}^{2}+\overline{Y_\alpha}^{2}\right),
\end{equation}
 which arises from the kinetic contribution of the energy density functional, carries two extra terms $\overline{X_\alpha}$ and $\overline{Y_\alpha}$, allowing us to manually set the coordinates of fixed vortices as well as their winding numbers. These terms are given by
\begin{equation}
\overline{X_{\alpha}} = \frac{n_{1,\alpha} x_{1,\alpha}}{r_{1,\alpha}^2}+ \frac{n_{2,\alpha} x_{2,\alpha}}{r_{2,\alpha}^2}, \quad\quad \overline{Y_{\alpha}} = \frac{n_{1,\alpha} y_{1,\alpha}}{r_{1,\alpha}^2}+ \frac{n_{2,\alpha} y_{2,\alpha}}{r_{2,\alpha}^2}. \nonumber \label{eqx}
\end{equation}

Finally, we remark that a similar procedure has already been sucessfully applied to superconductors in the framework of the Ginzburg-Landau theory  \cite{Chaves1,Chaves2}.

\section{Numerical Results}

The numerical solution of the CGP equations is obtained considering a two-dimensional rectangular system divided in a uniform square grid $4000\times2000$ with total dimension $1000\xi\times500\xi$, by using the finite-difference technique and a relaxation method suitable for non-linear differential equations. This leads us to the lowest-energy vortex structure that satisfies the constraint that vortices are placed in the fixed positions. The obtained condensate amplitudes are then substituted back in the energy density, which is numerically integrated, yielding the total energy $U=\int E(\psi_{\alpha})dxdy$ (dimensionless in units of $E_1=\mathcal{E}_1\xi^2$) of the corresponding vortex configuration. With the total energy as a function of the vortex-vortex distance at hand, our theoretical approach will then be used to investigate vortex-state solutions of the Gross-Pitaevskii equation in the presence of rotation, reflecting the real experimental conditions under which vortices are generated.

In what follows, we present the results first for BECs with contact interaction only, where we focus on the two-component case, and then also for coherently coupled two- and three-component BECs. For simplicity, we took the same number of particles per component for all situations, $N=5\times10^{4}$, as well as the intra-component couplings ($\gamma_{1}=\gamma_{2}=\gamma_{3}=1$). Since the investigation of the vortex-vortex interaction is performed within the miscibility condition: $\gamma_{1}\gamma_{2}-\gamma_{12}^{2}>0$, the used inter-component coupling parameters ​​are within the upper and lower limits $|\gamma_{12}|<1$. We denote by $U_{i , j, k}$ the total energy for the case of $i$ vortices placed in the first component, $j$ vortices placed in the second, and $k$ vortices placed in the third (if considered).

\subsection{Two-component BEC with contact interaction}

\begin{figure}[t]
\centerline{\includegraphics[width=1.00\linewidth]{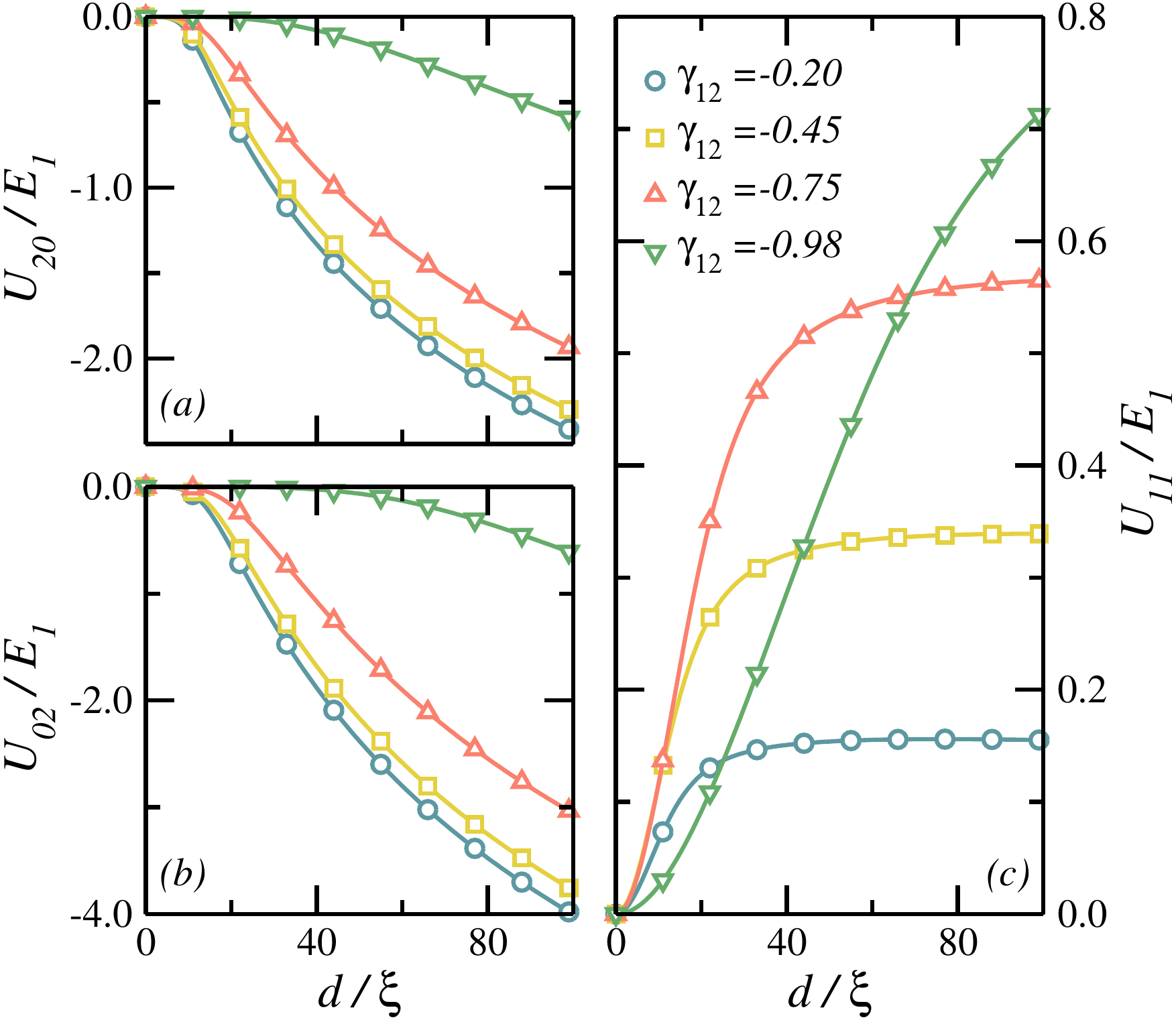}}
\caption{(color online) Vortex-vortex interaction potentials, considering $M_{12}=2.0$, for (a-b) two vortices in the same component  and (c) vortices in different components. Colours and symbols denote different inter-component coupling strength: $\gamma_{12} = -0.20$ (blue circles); $\gamma_{12} = -0.45$ (yellow squares); $\gamma_{12} = -0.75$ (red triangles); and $\gamma_{12} = -0.98$ (green downward triangles).} \label{Interactions}
\end{figure}

Let us first consider the case of a two-component condensate, with two vortices placed at $(x,y)=(\pm d/2,0)$ so that the distance between them is $d$. We shall consider components with different masses, as this has been found to lead to unusual lattice conformations in Ref. \cite{Kuopanportti}. By assuming a mass ratio of $M_{12}=2.0$ between the different components, we plot in Fig. \ref{Interactions} the vortex-vortex interaction for different values of the inter-component coupling strength: $\gamma_{12}=-0.20$ (circles); $\gamma_{12}=-0.45$ (squares); $\gamma_{12}=-0.75$ (triangles); and $\gamma_{12}=-0.98$ (downward triangles). Fig. \ref{Interactions} (a) and (b) represent the intra-component vortex-vortex interaction for the first and second component respectively, whereas Fig. \ref{Interactions}(c) depicts the inter-component vortex-vortex interaction.

It turns out that for a two-vortex configuration, the interaction potential is always a monotonic function of the distance, where the attractive or the repulsive behavior is determined by the properties of the particle-particle interaction. Nonetheless, one interesting feature can already be identified at this level: for \textit{inter}-component attraction, increasing the corresponding interaction strength $|\gamma_{12}|$ weakens the \textit{intra}-component vortex-vortex repulsion. This can be understood in terms of the form of the amplitudes $f_{} (x,y)$. Indeed, as the attractive \textit{inter}-component interaction becomes stronger, the energy is lowest for the largest possible superposition of the amplitudes, leading to a decrease in the density of one component in the spatial region where the other one has a vortex. Consequently, as $|\gamma_{12}|$ increases, these depletions should become more akin to vortices and the repulsive nature of the intra-component vortex-vortex interaction is softened by the depletion-vortex attractive interaction.

\begin{figure}[t]
\centerline{\includegraphics[width=1.0\linewidth]{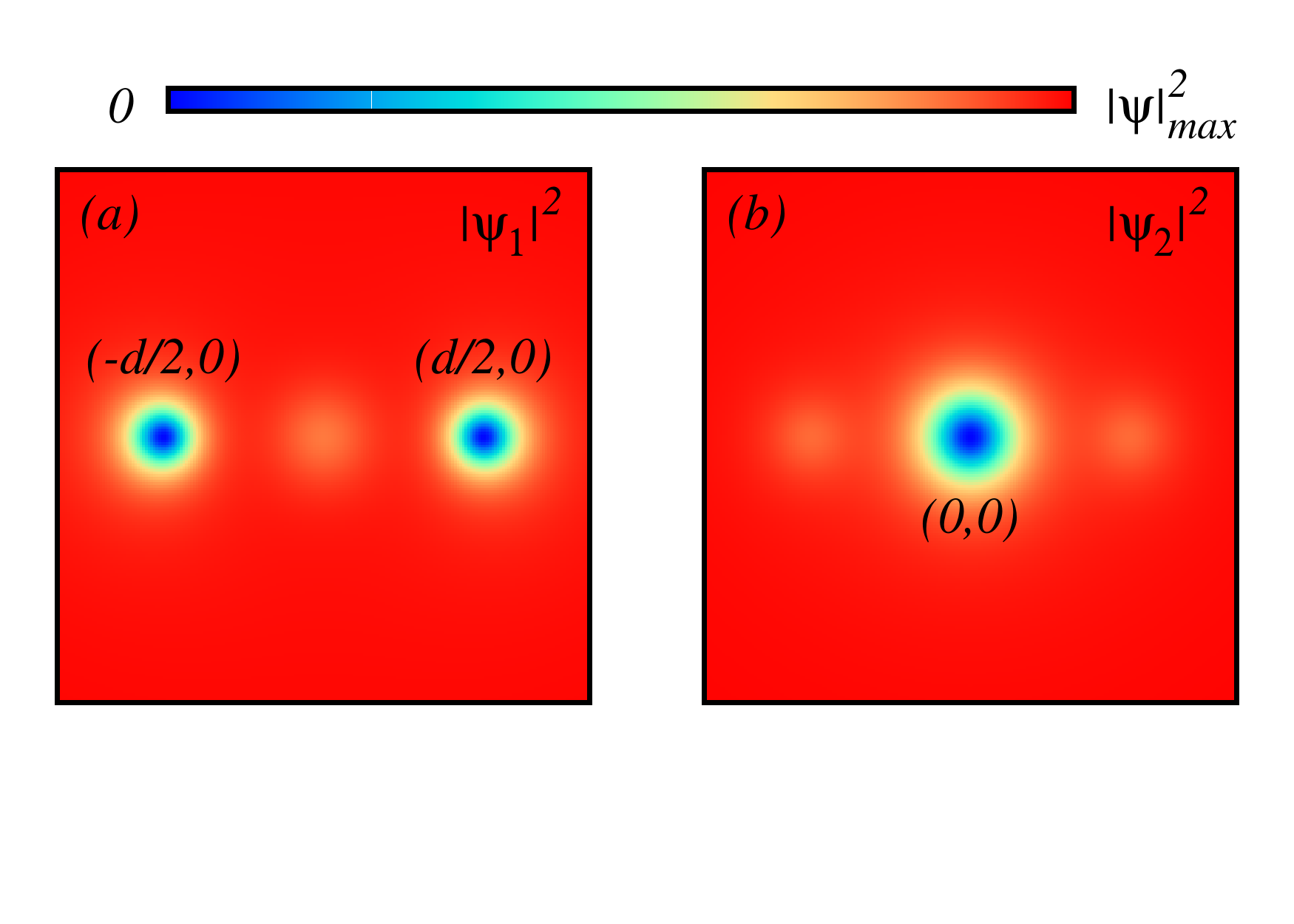}}
\caption{(color online) The occupation number density contour plots for: (a) first component with two vortices positioned at $(-d/2,0)$ and $(d/2,0)$ and (b) second component with a single vortex in the center of the mesh. Notice the depleted density in one component in the positions where the other component features a vortex. The investigation of the system energy with respect to $d$ for this conformation enables us to identify possible bound states in a two-component Bose-Einstein condensate with mass ratio $M_{12}=2.0$.} \label{Sketch1}
\end{figure}

Before we explore the consequences of the depletion induced in one component due to the presence of vortices in the other one, let us turn our attention to the mass ratio $M_{12}$. According to the Feynman relation of the vortex density in a rotating superfluid \cite{feynman}, the vortex density is proportional to the particle mass, according to
\begin{equation}
n_{v,\alpha}=\frac{m_{\alpha}\Omega_{R}}{\pi\hbar},
\end{equation}
where $\Omega_{R}$ is the angular rotation frequency of the condensate. Thus, for a mass ratio of $M_{12}=2.0$, it is natural to investigate situations where two vortices (one vortex) are placed in the heavier (lighter) component. To this end, we have pinned a vortex in the second component in the center of the mesh and have placed two further vortices in the first component, separated by a distance $d$, as illustrated in Fig. \ref{Sketch1} by a contour plot of both components. In Fig. \ref{Sketch1}, colors represent the density occupation number for both order parameters, and vortices coordinates are explicited as a function of the inter-vortex distance $d$.  In this case, the system energy is no longer a vortex-vortex interaction potential. However, this conformation should enable the identification of possible bound states provided by the competition between the inter-component and intra-component interactions, such as dimers and giant vortices \cite{VASchweigert,BXu,TCren,AAftalion,IDanaila,ALFetter,KKasamatsu,RGeurts} in the same component. Notice the weaker density in one component in the positions where the other component features a vortex, characterizing the depletion effect discussed before.

\begin{figure}[t]
\centerline{\includegraphics[width=1.0\linewidth]{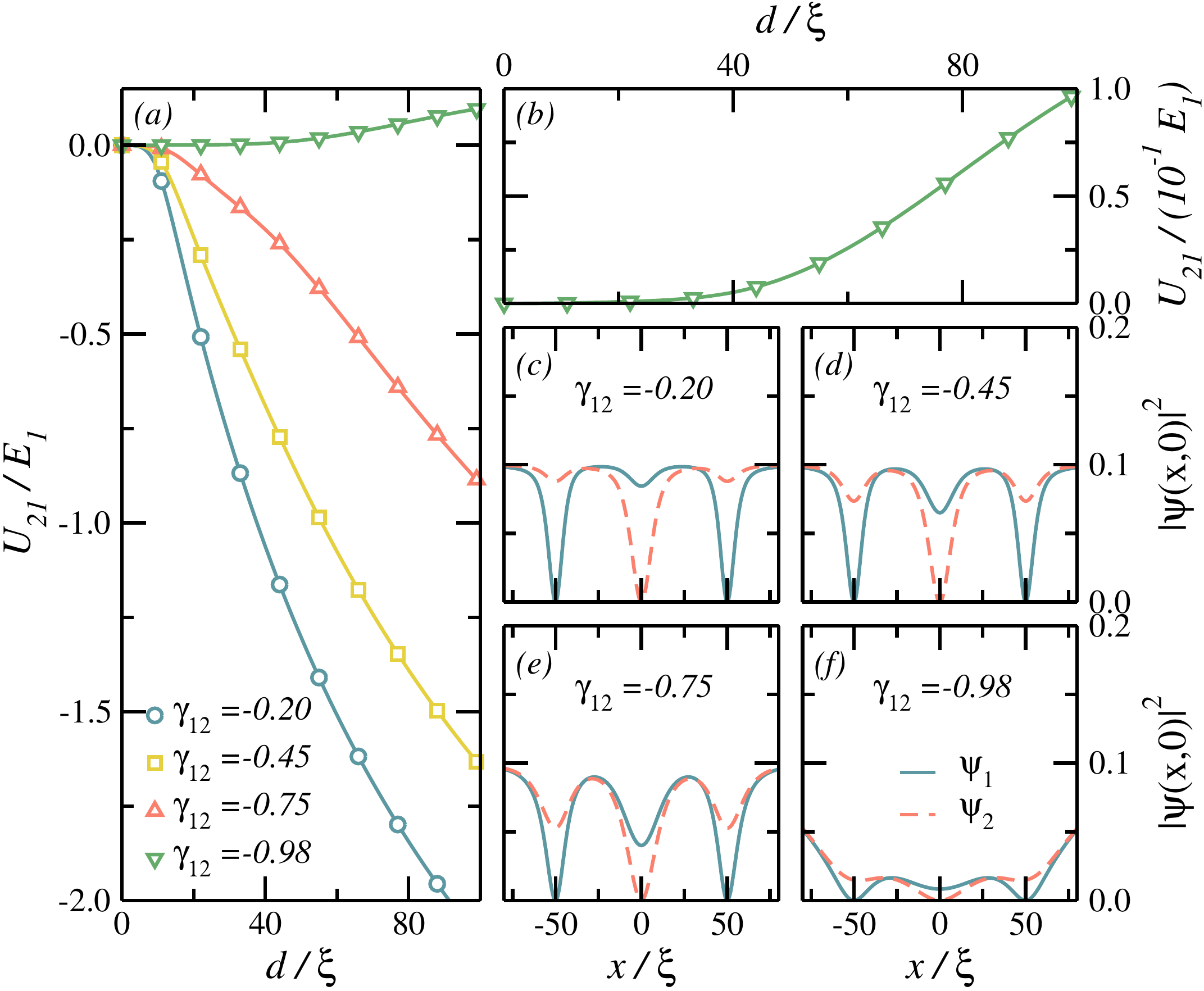}}
\caption{(color online) (a) Total energy for three vortices configuration $U_{21}$, characterizing a binary system with $M_{12}=2$. The vortex in the second component is fixed in the center of the mesh, whereas vortices in the first component are placed at symmetric positions separated by distance $d$. (b) A magnification of the particular case $\gamma_{12}$=-0.98 shows that in the favoured configuration, the two vortices in the first component are located on top of each other, and on top of the vortex in the second component. (c-f) Occupation number density of both components, considering vortices separated by 100 $\xi$, for all cases considered in (a), where the solid (dashed) line stands for the first (second) component.} \label{ComposedInteraction}
\end{figure}

The total energy of this three-vortex configuration is plotted in Fig. \ref{ComposedInteraction} (a) and (b), whereas Fig. \ref{ComposedInteraction}(c-f) demonstrates the previously discussed density depletion for several values of the relative interaction strength.

From Fig. \ref{ComposedInteraction}(a), one sees that for low values of $|\gamma_{12}|$ the energy is a monotonically decreasing function with the distance $d$, indicating repulsion between the vortices in the same component. However, this behavior is completely changed for $\gamma_{12}=-0.98$, where the total energy turns into a monotonically increasing function, allowing vortices of the same component to occupy the same position, thus, forming a giant vortex. Indeed, as one can see in the magnification of this result in Fig. \ref{ComposedInteraction}(b), the energy minimum corresponds to a vanishing distance between the vortices in the same component. At this point, it is evident that the depletion effect is the main cause for the behavioral change of the effective interaction. In fact, the strong coupling between the components makes the intra-component repulsive contribution less relevant than the inter-component vortex-vortex interaction, and the attractive effective interaction is achieved for every distance $d$. In Fig. \ref{ComposedInteraction}(c-f), the occupation number density for vortices separated by 100 $\xi$ shows the depletion effect, where depletions become more akin to vortices for higher values of inter-component particle coupling $|\gamma_{12}|$. It is noticed that vortices become larger when $\gamma_{12}$ is increased, causing the vortex strong overlap illustrated in Fig \ref{ComposedInteraction}(f). This effect depends on the inter-component particle coupling and vortex density. 

\begin{figure}[t]
\centerline{\includegraphics[width=1.0\linewidth]{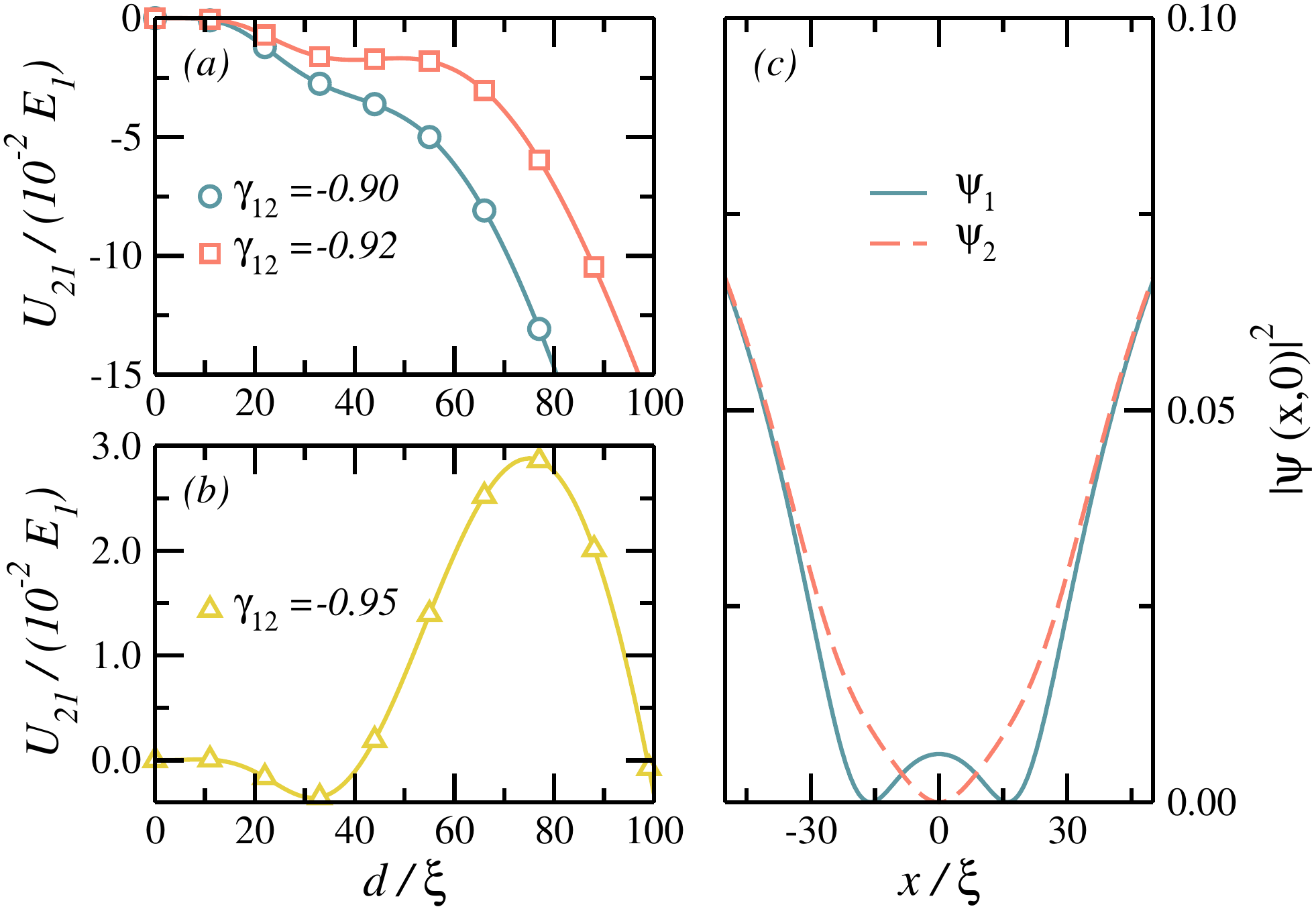}}
\caption{(color online) (a) Total energy of a three vortex structure in a two-component BEC, for $\gamma_{12}=-0.90$ (blue circles) and $-0.92$ (red squares). The curve bumps may lead to non-triangular lattices, despite the overall repulsive behavior. (b) Particular case of a triple-vortex structure with $\gamma_{12}=-0.95$. The minimum far from the origin allows for the formation of dimers for specific vortex densities. (c) Dimer configuration profile for $\gamma_{12}=-0.95$.} \label{EspecialCases}
\end{figure}

Before going into further details concerning vortex core, let us first analyze how the effective interaction between vortices changes from repulsive to attractive, by considering intermediate values of $\gamma_{12}$ between $\gamma_{12}=-0.75$ and $\gamma_{12}=-0.98$. We have found that, for $\gamma_{12}=-0.90$ and $\gamma_{12}=-0.92$, as illustrated in Fig. \ref{EspecialCases}(a), the energy curves still have repulsive characters, but they exhibit shoulders at finite distances, which could indicate the formation of non-triangular lattices in these minima of the interaction potential, despite the overall repulsive behavior. In Fig. \ref{EspecialCases}(b), by setting $\gamma_{12}=-0.95$, we have obtained a total energy which exhibits a non-monotonic behavior and acquires a clear minimum at finite inter-vortex distance. This strongly indicates that the presence of a vortex in one component can cause the formation of a bound state of two vortices (dimer) in the heavier component, illustrated in Fig. \ref{EspecialCases}(c). This results corroborates the states obtained in numerical experiments of Ref. \cite{Kuopanportti}, where the Abrikosov lattice of vortex dimers in one component, sitting on single vortices in the other, was found in a binary mixture.

\begin{figure}[t]
\centerline{\includegraphics[width=1.0\linewidth]{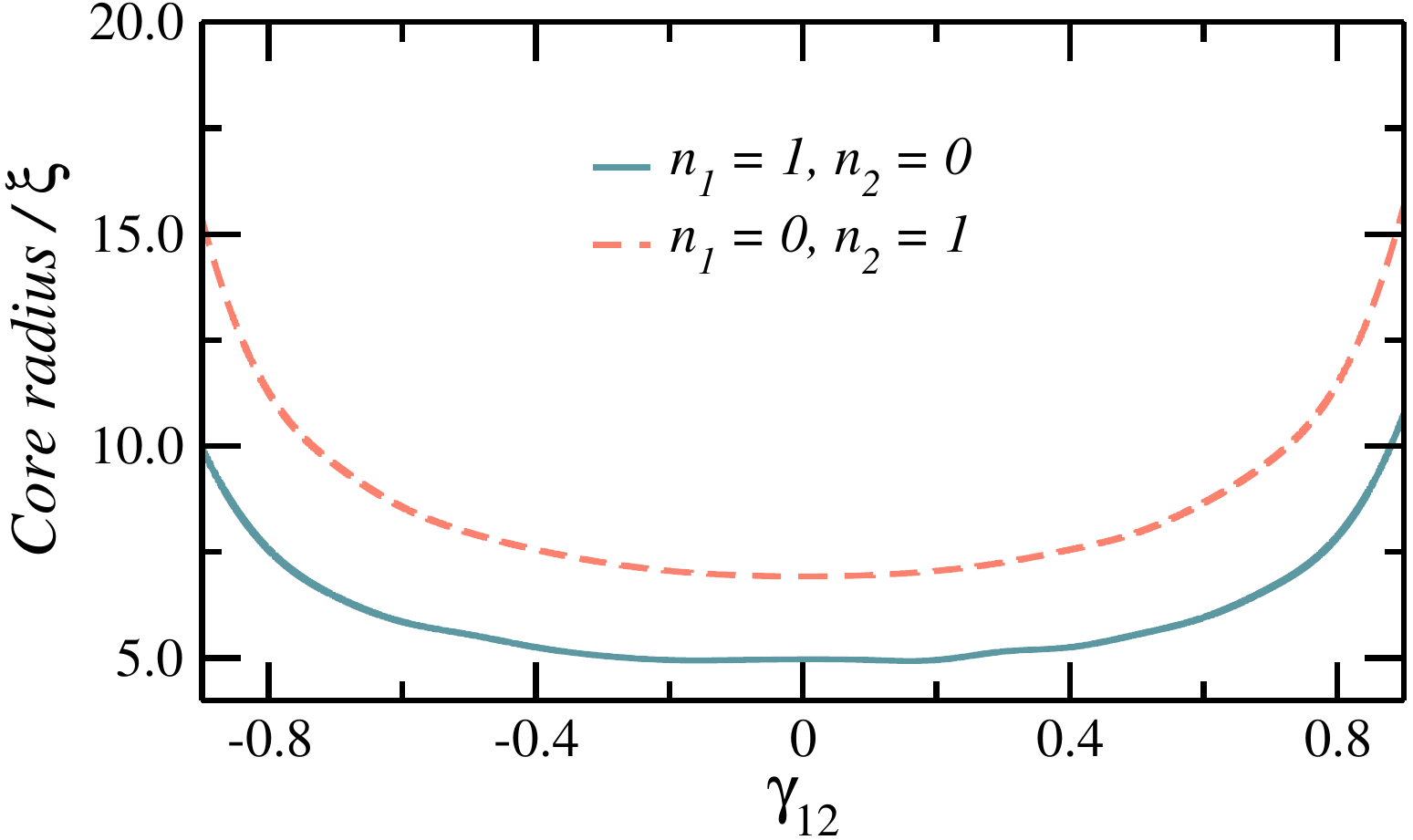}}
\caption{(color online) Vortex core size as a function of the inter-component coupling, for $M_{12}=2.0$. The blue solid line (red dashed line) represents the core radius of a vortex in the first (second) component.} \label{CoreSize}
\end{figure}

In addition to effects caused in the vortex-vortex interactions, leading to different lattice conformations, the depletion effect also leads to formation of vortices with different core sizes according to the choice of the $\gamma_{12}$ parameter. Actually, this was pointed out before in Fig. \ref{ComposedInteraction}(c-f), where more negative values of $\gamma_{12}$ result in vortices with larger core sizes. To investigate this dependence, in Fig. \ref{CoreSize}, by setting a vortex in only one of the two components, the vortex core radius was measured for different values ​​of inter-component coupling, between $\gamma_{12}=-0.90$ and  $\gamma_{12}=0.90$. This was done by measuring the distance between the center of the vortex and the point at which the order parameter decreases by half of its long-range convergence value.

Since the components have different mass, in Fig. \ref{CoreSize} we investigated this effect for vortices in both components, where the core size for a first(second)-component vortex is presented by the solid (dashed) line. It is observed that the more the two components are coupled, the larger the vortex radii will be. This effect actually comes from the depletion originated by the vortex depletion which contributes to an increased vortex size. For both components, this behavior is qualitatively the same, suggesting that the different mass causes simply a shift between two curves.

Let us at this point address the question of how realistic the present predictions are from an experimental point of view. For example, a two species BEC with a tunable interspecies interaction has already been produced from the mixture of $^{87}$Rb and $^{41}$K, for which one has $M_{12}\approx2.1$ \cite{thalhammer}. In that particular study, a Feshbach resonance around a magnetic field of $B\approx79$G allows for tuning $a_{12}$ from positive to negative values. For example, for $80.7$G $a_{12}\approx-185a_{0}$ is reported, with $a_{0}$ the Bohr radius. This clearly opens up the possibility of obtaining the appropriate negative values of $\gamma_{12}$ needed to experimentally observe the predicted vortex states. We also remark that, for a $^{87}$Rb-$^{85}$Rb mixture, tunability of the interspecies interaction has already been used to probe various mean-field regimes such as spatial separation as well as the formation of long lived droplets (see Ref. \cite{papp}).

\subsection{Vortex dimers and trimers in coherently coupled BECs}

In this subsection, we consider the presence of an internal coherent coupling of the Rabi type between the different components. In order to do so, an extra term should be added to the energy density functional, as pointed out in Sec. \ref{SecI}. For a system with $N_{c}$ components, the Rabi contribution in Eq. (\ref{Eq2}) can be rewritten in the form 
\begin{equation}
-2\sum_{i}^{N_{c}}\sum_{j>i}^{N_{c}}\omega_{ij}|\psi_{i}||\psi_{j}|\cos(\theta_{i}-\theta_{j}).
\nonumber
\end{equation}
If the signs of $\omega_{ij}$ coefficients are all positive, the ground state energy is achieved when all the phases $\theta_{i}$ are the same. Recalling that the phases fix the position of the vortices, this means that vortices of different components would overlap, featuring an attractive potential. On the other hand, by setting positive values for the contact coupling $\gamma_{ij}$, a repulsive inter-component vortex-vortex interaction emerges. The balance between these two types of interaction can lead to molecular conformations of vortices. Unlike the previous case, these bound states do not arise due to the mass difference between the components, but due to the competing interactions induced by the Rabi contribution. To illustrate the above argument, consider, for instance, the case with two components with equal masses $M_{1\alpha}=1$, with one vortex in each component.

\begin{figure}[t]
\centerline{\includegraphics[width=1.0\linewidth]{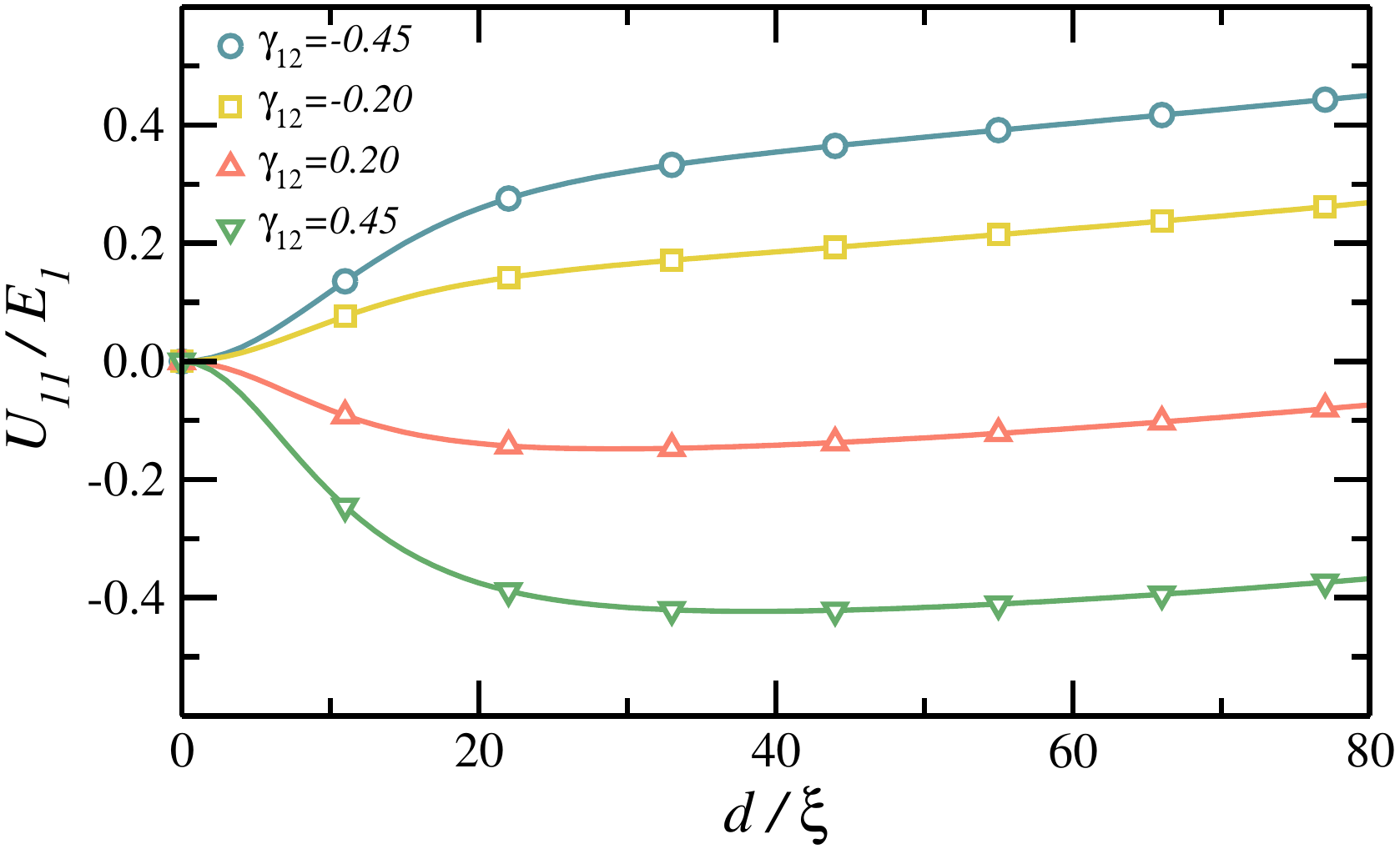}}
\caption{(color online) Two-vortex inter-component interaction potential in the case of competing contact and Rabi coupling, $\omega=2.1\times10^{-5}$ and $M_{12}=1$, for several values of $\gamma_{12}$.} \label{RabiEffect}
\end{figure}

 As illustrated in Fig. \ref{RabiEffect}, a vortex dimer may be formed in the case of a short range repulsive and long range atractive inter-component potential, as observed for positive values of $\gamma_{ij}$, where a minimum of interaction energy is found at finite inter-vortex distance due to a competition between two types of inter-component interaction. On the other hand, negative values of $\gamma_{12}$ lead to an extended attractive interaction in the long range. 

\begin{figure}[t]
\centerline{\includegraphics[width=1.0\linewidth]{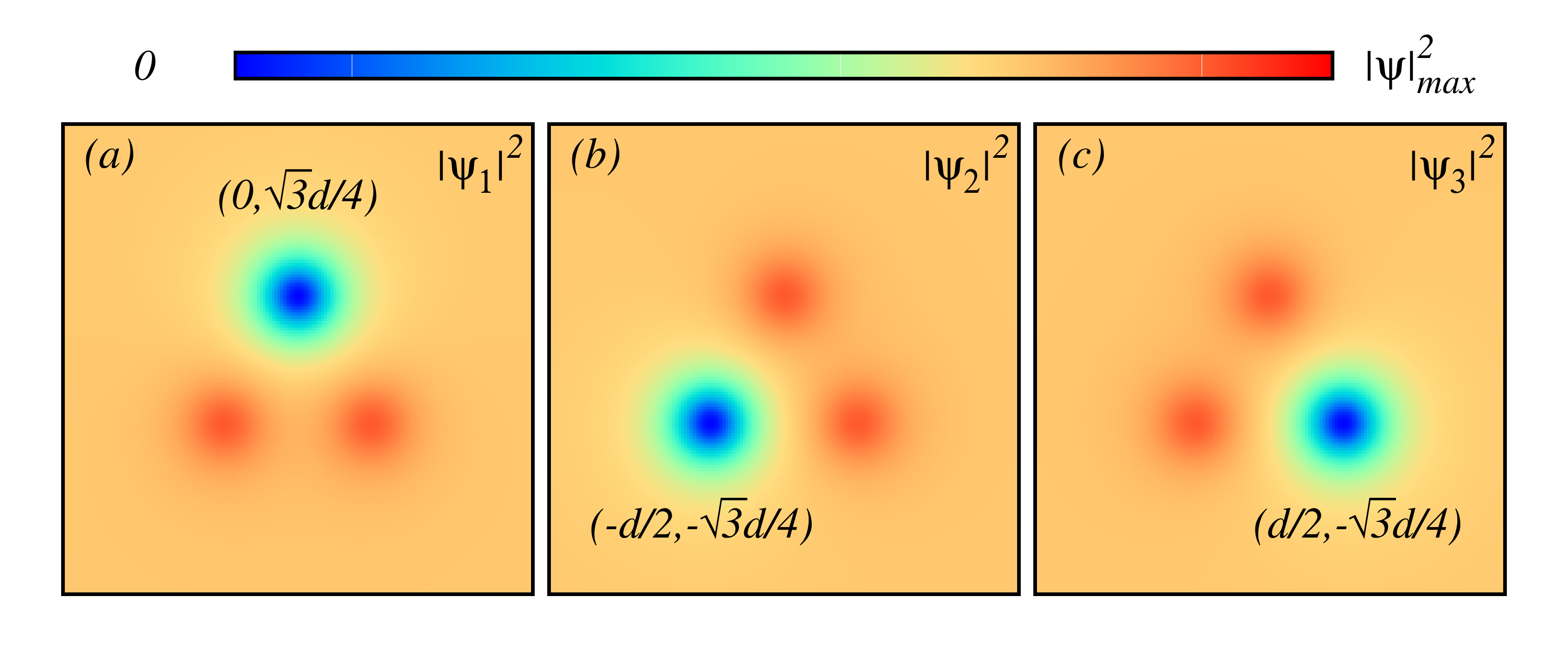}}
\caption{(color online) The occupation number density contour plots of: (a) first component with a vortex positioned at $(0,\sqrt{3}d/4)$; (b) second component with a vortex positioned at $(-d/2,-\sqrt{3}d/4)$ and ; (c) third component with a vortex positioned at $(d/2,-\sqrt{3}d/4)$.} \label{Sketch2}
\end{figure}

Let us now address the question of the formation of vortex trimers in a three-component condensate, i.e., bound states consisting of three vortices. In an attempt to reduce the excessive number of parameters, we shall consider the case with $\omega_{12}=\omega_{13}=\omega_{23}=\omega$ and $\gamma_{12}=\gamma_{13}=\gamma_{23}=\gamma$. This choice also helps in finding the minimum energy configuration, since in this case, for symmetry reasons, it is expected that the three vortices are arranged equidistantly. Thus, we manually placed the vortices at the vertices of an equilateral triangle of side $d$, whose loci then read $S_{1}=(0,\sqrt{3}d/4)$, $S_{2}=(-d/2,-\sqrt{3}d/4)$ and $S_{3}=(d/2,-\sqrt{3}d/4)$. This is illustrated in Fig. \ref{Sketch2}, where contour plots of the densities of the three components are shown. Notice the presence of the depletion effect in this case as well. Subsequently, we investigated the minimum energy conformation as a function of $d$ for several values of $\omega$.

\begin{figure}[t]
\centerline{\includegraphics[width=1.0\linewidth]{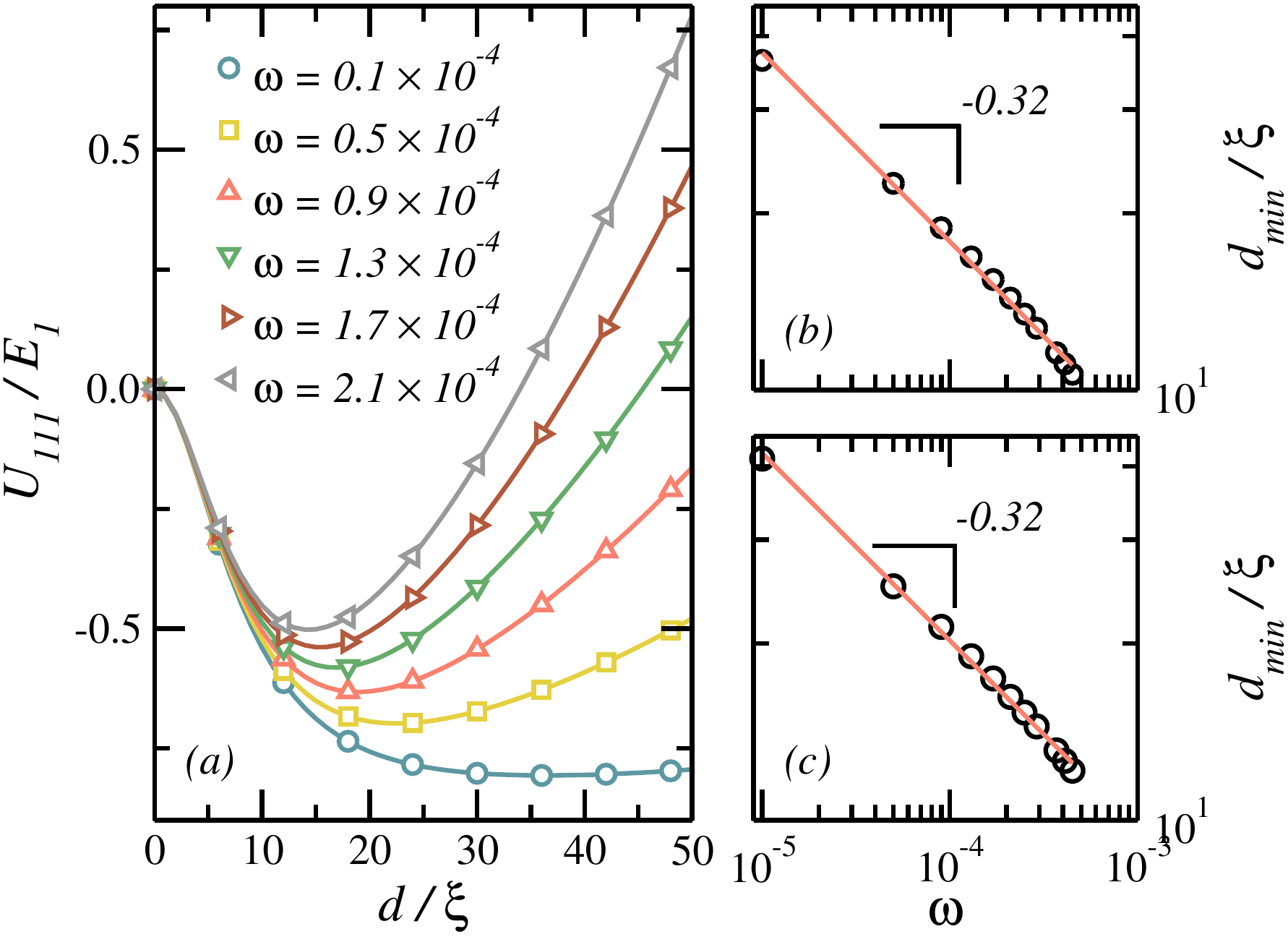}}
\caption{(color online) (a) Total energy $U_{111}$, for the equilateral triangle vortex configuration as a function of the side $d$ for $\gamma=0.45$ and for several values of $\omega$. (b) Plot of the optimal side of the equilateral vortex triangle against $\omega$ for $\gamma=0.45$. The red line represents a power law with exponent $-0.3214$ and coefficient $0.9265$. (c) Same as (b) but for $\gamma=0.60$. The red line represents a power law with exponent $-0.3186$ and coefficient $1.0746$.  } \label{Molecule}
\end{figure}

In Fig. \ref{Molecule}(a), where we considered $\gamma=0.45$, there is indeed a finite value $d_{min}$ of the triangle side that minimizes the interaction energy $U_{111}$. Moreover, Fig. \ref{Molecule} (b) and (c) display power law dependencies of $d_{min}$ in $\omega$, for $\gamma=0.45$ and $\gamma=0.60$, respectively, which explains the vortex trimer configurations observed e.g. in Fig. 3 of Ref. \cite{Eto}. 
Despite the fact that the exponents are quite similar in the two considered cases, they should depend on both $\gamma$ and number of particles $N$. Indeed, considering the extreme case where $\gamma$ vanishes, the inflection point of the vortex interaction should always be at $d=0$ for any value of $\omega$, leading to a vanishing exponent in the power law dependence $d_{min}(\gamma)$.

\begin{figure}[t]
\centerline{\includegraphics[width=1.0\linewidth]{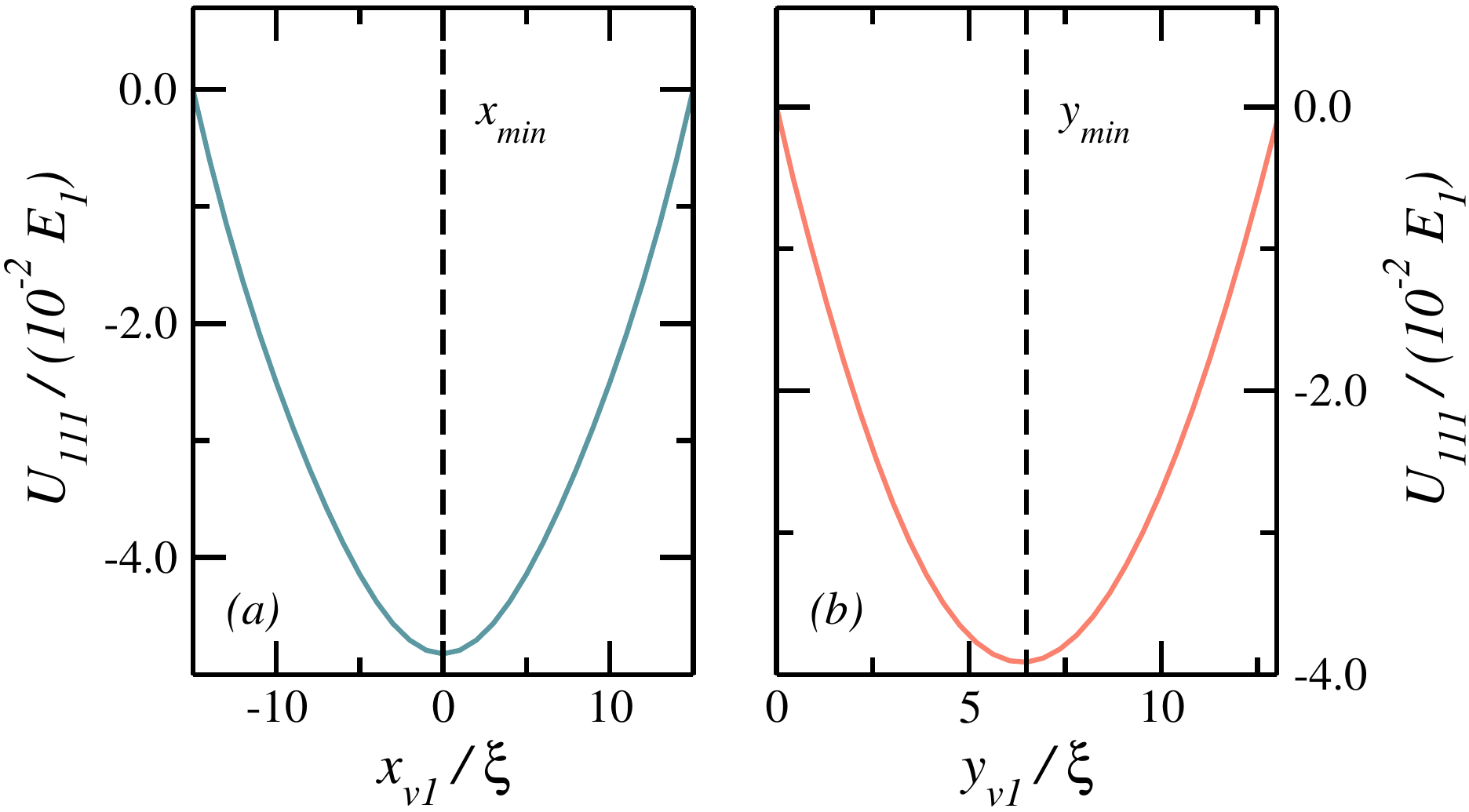}}
\caption{(color online) (a) Total energy $U_{111}$ as a function of the displacement of the first component vortex $x_{v1}$ from $S_{1}$ position in the $x$ direction. (b) Total energy $U_{111}$ as a function of the displacement of the first component vortex $y_{v1}$ from the origin in the $y$ direction. For both cases, we have considered $\gamma=0.45$ and $\omega=1.7\times10^{-4}$. The vertical dashed line represents the coordinates of first component vortex for the minimum energy equilateral triangle configuration. }\label{GlobalMinimum}
\end{figure}

Since the geometry was imposed from the beginning by setting the vortices at the vertices of the equilateral triangle, the existence of a value $d_{min}$ which minimizes the total energy does not guarantee that this configuration corresponds to the ground state. In order to allow for different configurations and thereby check if the equilateral triangle is really a minimum of the energy, we have allowed for different positions of the vertex $S_{1}$, while keeping $S_{2}$ and $S_{3}$ fixed for $\omega=1.7\times10^{-4}$ and $\gamma=0.45$. As can be seen from Fig. \ref{GlobalMinimum}, the equilateral triangle configuration turns out to be stable, corresponding to at least a local minimum of the interaction energy.

Although not very common, there are other ways to obtain bound vortex states which do not arise from a competition between different types of interaction. In fact, it is also possible to obtain dimers and trimers exclusively from Rabi coupling in condensates with at least three components. These vortex states arise from frustration between the phase locking tendencies. As it was mentioned previously, a set of positive Rabi frequencies $\omega_{12}=\omega_{13}=\omega_{23}=\omega>0$ would lead to $\theta_{1}=\theta_{2}=\theta_{3}$ as a minimum of the energy, causing the same spatial occupation for all three vortices. However, by assuming, for example, $\omega_{12}=\omega_{13}=-\omega_{23}=\omega$, it is not possible to satisfy the minimization energy condition $\cos(\phi_{1}-\phi_{2})=\cos(\phi_{1}-\phi_{3})=1$ and $\cos(\phi_{2}-\phi_{3})=-1$ simultaneously, which characterizes a frustrated system \cite{EBabaev,VStanev,XHu,NVOrlova}. 

Since the frustration phenomenon arises basically from the problem of minimizing the total energy with respect to the order parameter phases, we must redefine our ansatz, adding extra phases to our order parameters in order to account for the phase tendencies which minimize the energy. Thus, by assuming an ansatz of the form $\psi_{n}(x,y)=\sqrt{N}e^{i\theta_{n}}e^{-i\phi_{n}}f_{n}(x,y)$, the consequences of choosing Rabi frequencies which lead to a frustrated system were investigated for $\omega_{12}=\omega_{13}=-\omega_{23}=\omega$ and $\gamma_{12}=\gamma_{13}=\gamma_{23}=0$, where we have considered $\phi_{1}=\phi_{2}=0$ and several values for $\phi_{3}$.

\begin{figure}[t]
\centerline{\includegraphics[width=1.0\linewidth]{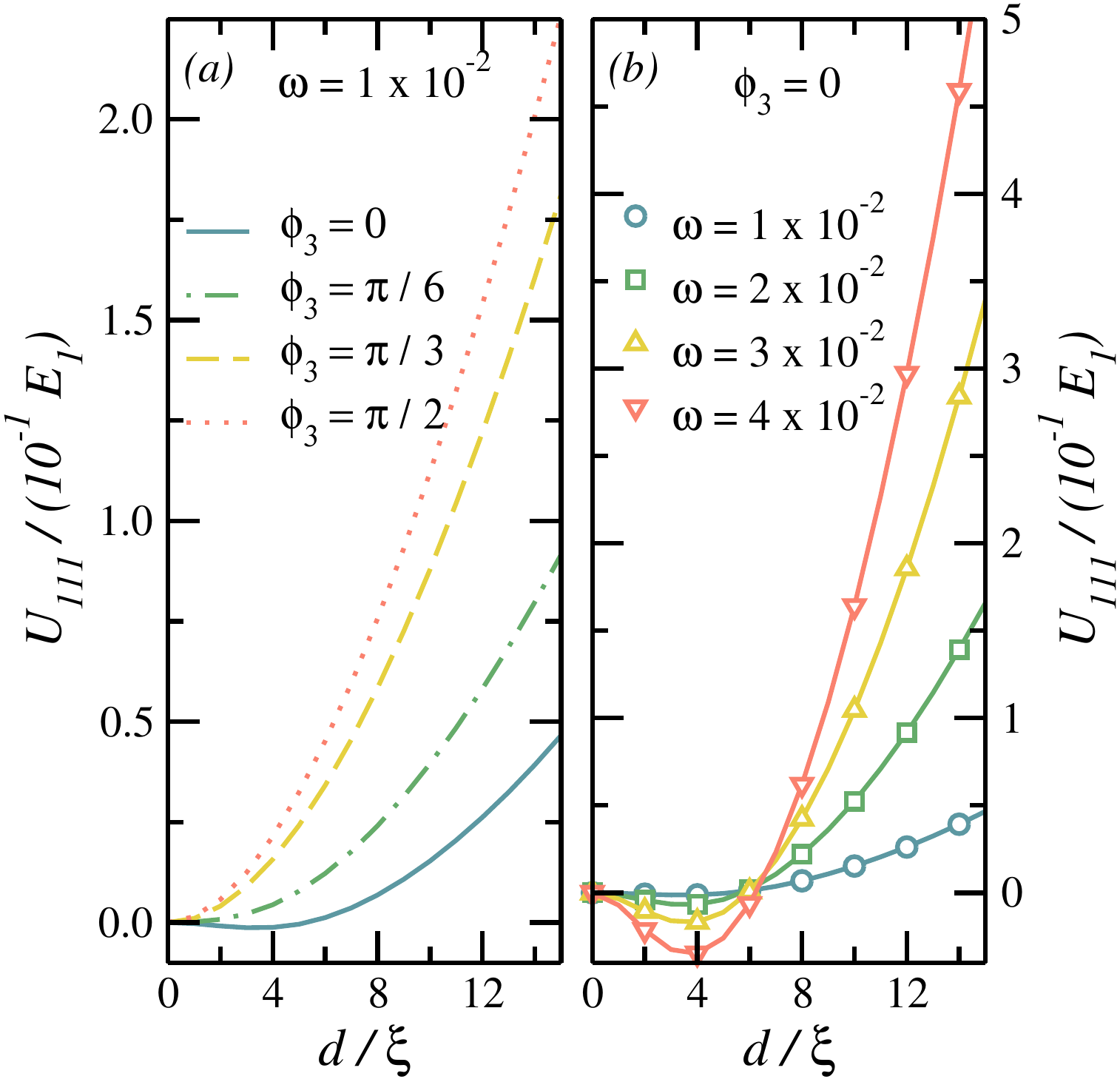}}
\caption{(color online) (a) Total energy $U_{111}$, with vortices of the first and second components pinned in the center of the mesh, as a function of the third component vortex distance $d$ from the origin, for different values of the third component extra phase $\phi_{3}$: 0 (solid blue line), $\pi/6$ (green dashed dotted line), $\pi/3$ (yellow dashed line) and $\pi/2$ (red dotted line), with $\omega_{12} =\omega_{13}=-\omega_{23}=1\times10^{-2}$. (b)  Total energy $U_{111}$ for the same conformation of (a), keeping $\phi_{3}=0$ and assuming different values of $\omega$: $\omega=1\times10^{-2}$ (blue circles), $\omega=2\times10^{-2}$ (green squares), $\omega=3\times10^{-2}$ (yellow upward triangles) and $\omega=4\times10^{-2}$ (red downward triangles).}\label{Frustration}
\end{figure}

In Fig. \ref{Frustration}(a), we pinned vortices of the first and second components on top of each other, in the center of the mesh and displaced the vortex of the third component at distance $d$ from the origin. This procedure was made for different values of phase $\phi_{3}$. The obtained results show a potential that suggests the formation of a bound vortex state, with the third component vortex outside the origin, for $\phi_{3}=0$, where the potential minimum is at $d\approx4\xi$. This conformation arises from the competition between the attraction with the vortex of the first component and the repulsion with the vortex of the second component. By symmetry, the same result should also appear for $\phi=\pi$. This configuration has the lowest energy among the other investigated cases, however, we can not guarantee that this conformation is the ground state. Actually, finding the ground state conformation would require the variation of all three extra phases and also other vortices positions in the grid, which is very expensive computationally and is left for future investigation. Nevertheless, these results are sufficient to ensure that vortex molecules are stable states. In Fig. \ref{Frustration}(b), we show that increasing the Rabi coupling makes the frustration effect even more robust and leads to a deeper minimum in the interaction potential.

In view of the present predictions concerning Rabi coupled BECs, it becomes important to consider the possibilities of experimental realization of adequate samples. In that respect, we focus on the phenomena of interest, namely Rabi coupling and homogeneous confinement. They have both already been experimentally achieved with $^{87}$Rb. Before we turn to the particular experimental setups, we recall that in our calculations the parameter governing the Rabi coupling is $\omega = w\bar{\rho}_1/{\mathcal E}_1$, where $w=\hbar\Omega$ relates the Rabi energy $w$ to the Rabi oscillation frequency $\Omega$.

We first consider a recent study in Ref. \cite{chomaz}, demonstrating long-range coherence in quasi-two-dimensional Bose gases. There, two-dimensional densities of about a few hundred $^{87}$Rb-atoms per square micrometer are achieved and combined with harmonic trapping in the third direction whose frequency $\nu_z$ ranges from $350$ up to $1500$ Hz. Moreover, for $^{87}$Rb, the (three-dimensional) s-wave scattering length is $a_s \approx 100a_{0}$, with $a_{0}$ the Bohr radius. For the values of the Rabi couplings we turn to Ref. \cite{Matthews}, where Rabi oscillation frequencies of the order $2\pi\times10^{2}s^{-1}$ have been realized. We estimate the Rabi coupling parameter $\omega$ by calculating $E_1=\xi^{2}{\mathcal E}_{1}$ for these values of the particle density and Rabi frequencies, to arrive at $\omega=\frac{2m_{1}\Omega}{\hbar\bar{\rho}_{1}}=\frac{2}{a_{\textrm Rabi}^2\bar{\rho}_1} \approx 1.7\times 10^{-2}$, where we have associated a length $a_{\textrm Rabi}=\sqrt{\hbar/m_{1}\Omega}$ with the Rabi oscillation. This value is already higher than ones needed for the effects discussed in this manuscript.

It is also possible to improve the estimates of the actual experimental values of the coherent coupling by including the effect of the transversal harmonic trapping, rendering the calculation more realistic. Indeed, in order to obtain typical values of $\omega$ in quasi-two-dimensional systems, one should correct the s-wave scattering parameter $g$ for the freezing of the third dimension. This is done by dividing the three-dimensional $g_{3D} = 4\pi \hbar^2 a_s/m_{1}$ by $\sqrt{2\pi}a_z$, with $a_z$ being the oscillator length in the third dimension \cite{fischer}. If one then associates the energy $E_1$ with the contact interaction energy $g\bar{\rho}_{1}$, a slightly different expression for the coherence parameter emerges $\omega=\frac{1}{a_{\textrm Rabi}^2\bar{\rho}_{1}}\frac{a_{z}}{\sqrt{8\pi}a_s}\approx 1\times 10^{-1}$. Notice that, besides the fact that this value of $\omega$ is even higher in quasi-two-dimensional traps, this expression allows to identify the trap frequency in the third dimension, in addition to the Rabi oscillation frequency and the particle density, as tuning knobs for studying the effects of coherent coupling in multi-component BECs.

\section{Conclusion}

In summary, we have presented a method for calculating pairwise vortex-vortex interaction potentials in multi-component Bose-Einstein condensates, capable of revealing the underlying reasons for the unusual vortex configurations, such as different lattice geometries and few-vortex clusters. We have applied our theory in specific examples and could thereby clarify the formation of a lattice of dimers in rotating two-component BECs as well as the dimer/trimer state in a coherently coupled two/three-component BEC by pointing out the role of the non-monotonic interaction potential with respect to the inter-vortex distance in both cases. Our analysis of the present day experimental capabilities, combined with simple estimates of relevant parameters, indicates that considered effects are on the verge of being directly observed in the lab. We remark that the present theory can be straightforwardly adapted to any number of condensate components, providing a tool to anticipate different aspects of vortex physics in Bose-Einstein condensates. As a matter of fact, even for one-component BECs featuring the anisotropic and long-range dipolar interaction, vortex phases such as bubbles and stripes have been foreseen \cite{cooper_stripes,zhang_stripes,kishor}, for which a thorough understanding of the role of vortex-vortex interaction is still lacking.

\section{Acknowledgment}
This work was supported by the National Council for Scientific and Technological Development (CNPq-Brazil), the Coordination for the Improvement of Higher Education Personnel (CAPES-Brazil), Research Foundation Flanders (FWO), and the bilateral FWO-CNPq program between Flanders and Brazil. M.V.M. acknowledges support from CAPES-PVE program (BEX1392/11-5).

\end{document}